\documentclass{article}

\PassOptionsToPackage{numbers,compress}{natbib}
\usepackage[main, final]{neurips_2026_vericode}

\usepackage[utf8]{inputenc} 
\usepackage[T1]{fontenc}    
\usepackage{hyperref}       
\usepackage{url}            
\usepackage{booktabs}       
\usepackage{amsfonts}       
\usepackage{nicefrac}       
\usepackage{microtype}      
\usepackage[table]{xcolor}  
\usepackage{graphicx}
\usepackage{placeins} 

\usepackage{fvextra}
\usepackage{float}
\input{code/highlighting_style.tex}
\fvset{extra=true}
\VerbatimPygments{\PYG}{\PYG}
\usepackage{tikz}
\usepackage{subcaption}

\newlength{\cellw}
\newlength{\cellh}
\newcommand{\cell}[3]{%
  \begin{tikzpicture}[baseline=-0.5ex]
    \ifnum#1=1 \fill[black] (0,0) rectangle (\cellw,\cellh);\fi
    \ifnum#2=1 \fill[black] (\cellw,0) rectangle (2\cellw,\cellh);\fi
    \draw[gray,very thin] (0,0) rectangle (2\cellw,\cellh);
    \draw[gray,very thin] (\cellw,0) -- (\cellw,\cellh);
    \ifnum#3=1
      \fill[white,draw=black,very thin]
        (\cellw,0.5\cellh) circle (0.22\cellh);
    \fi
  \end{tikzpicture}%
}
 
\title{Proving LLM-Generated GPU Kernels Correct and Safe: Floating-Point Verification in Lean and VerCors}
\workshoptitle{AI for Verifiable Coding}

\author{%
  Archibald Browne\thanks{Equal contribution.} \\
  Imperial College London \\
  Logos Research \\
  \texttt{ajb421@ic.ac.uk}
  \And
  Eugene Lee\footnotemark[1] \\
  Imperial College London \\
  Logos Research \\
  \texttt{el625@ic.ac.uk}
  \And
  Cristopher Salvi \\
  Imperial College London \\
  Logos Research \\
  \texttt{c.salvi@logosresearch.ai}
}

\begin{document}

\maketitle

\begin{abstract}
Modern statistical model development begins with a high-level specification in an automatic differentiation framework such as PyTorch, JAX, or TensorFlow, which is then lowered to a target backend by a compiler or implemented manually by a research engineer. In this paper, we study whether frontier language models can correctly perform the latter task by generating GPU kernels from high-level benchmark specifications. We introduce a framework in which a language-model agent uses Velvet, a program verifier for imperative programs embedded in Lean, to prove semantic equivalence between a high-level specification and a generated CUDA kernel under an explicit set of assumptions. It also asks the LLM to prove thread safety and data race freedom of the generated kernel in VerCors, a deductive verifier for parallel code. Our main contribution to the formal verification of LLM-generated GPU kernels is showing that a combination of Lean automation and language-model-guided proof construction can verify a broad set of basic neural-network building blocks over floating-point values, allowing LLMs to implement semantically correct and safe kernels. In our framework, floating-point verifications never rely on compiled code and are always checked via the Lean kernel.

\end{abstract}

\section{Introduction}
\label{sec:Introduction}



Modern ML model development begins with a high-level specification in frameworks such as PyTorch, which is lowered to hardware-specific kernels by compiler backends or expert engineers. This translation can leave substantial performance unrealised until the software stack catches up: on the same 512 H100 GPUs, NVIDIA reported a 27\% year-over-year improvement on MLPerf GPT-3 training from software optimisations alone \cite{eassa2024mlperf}. 

KernelBench tasks language models with automating this process by generating optimised GPU kernels directly from PyTorch specifications \cite{ouyang2025kernelbench}. However, KernelBench-Verified shows that LLM-generated kernels can exploit weaknesses in finite correctness tests, producing implementations that appear correct and fast under the benchmark distribution but fail under broader inputs \cite{zhang2026kernelbenchverified}. This motivates a verification layer in which the model must make its assumptions explicit and prove that the generated kernel is semantically equivalent to the high-level specification, rather than relying on its own tests or informal reasoning about correctness.

To tackle this problem, we implement an agentic framework making use of a formal verification layer in order to mechanically check semantic equivalence between high-level specifications and generated GPU kernels, as well as verifying the data race freedom of the parallel code. Our contributions are as follows: \begin{enumerate} \item We demonstrate that Lean automation combined with language-model-guided proof construction can establish semantic equivalence for a broad set of basic neural-network kernels over floating-point values under explicit assumptions. \item We identify semantic failures in LLM-generated GPU kernels that pass conventional correctness tests and verification approaches that reason over $\mathbb{R}$ or $\mathbb{Z}$. \end{enumerate}

In Section~\ref{sec:Background}, we briefly review related work in the area. Formal verification of LLM-generated code is a rapidly growing area and ensuring that the formal claims match up with the generated code is a key bottleneck, especially when floating point arithmetic is involved or highly important to the application of the generated code. In Section~\ref{sec:ArchitectureAndSetup}, we move on to describe the architecture of our agentic pipeline. The key components are:
\begin{itemize}
    \item Deterministic compilers from both PyTorch specifications and a limited set of CUDA to Lean;
    \item Generator and Prover agents which are designed for creating the resulting CUDA code and proving its semantic equivalence with the PyTorch specification respectively;
    \item The leveraging of Velvet \cite{gladshtein2026velvet} in order to reason over imperative code in Lean via a Dafny-like manner;
    \item The VerCors \cite{Armborst_The_VerCors_Verifier_2024} gate which aims to provide thread and memory safety of the generated code.
\end{itemize}
We also detail the purpose of each of the technologies we employ in our pipeline, and how they assist with our goal of providing a framework for producing verified GPU code. In Section~\ref{sec:Results}, we experiment with setting the underlying model of the framework to both GPT-5.6 Terra and Claude Opus 5,\footnote{Both the \emph{fast} variants, which simply allow for increased throughput rather than affecting model abilities.} and review some statistics about both runs. In Section~\ref{sec:CaseStudy}, we provide representative case studies demonstrating examples in which the harness was able to catch semantic issues. Further case studies are provided in Appendix~\ref{app:casestudies}. We finally conclude in Section~\ref{sec:Conclusion} with a summary of our findings and a discussion of the limitations of our approach.


\section{Background}
\label{sec:Background}





\paragraph{LLM-generated GPU kernels.}
KernelBench introduced a benchmark for generating optimised GPU kernels from PyTorch specifications, evaluating both correctness and speed \cite{ouyang2025kernelbench}. KernelBench-Verified later showed that finite correctness tests can be exploited by generated kernels, producing implementations that appear correct under the benchmark distribution but fail under broader inputs \cite{zhang2026kernelbenchverified}. These results motivate stronger correctness guarantees than empirical testing alone.

\paragraph{Formal verification of GPU kernels.}
Classical GPU verification has focused largely on safety and concurrency properties. GPUVerify checks CUDA and OpenCL kernels for data races and barrier divergence \cite{betts2012gpuverify}, while VerCors supports deductive verification of CUDA programs using contracts and permissions, including functional properties of parallel algorithms \cite{safari2022vercorscuda}. Formal models of PTX memory behaviour further provide a semantic basis for reasoning about low-level GPU execution \cite{lustig2019ptx}. More recent systems move toward semantic equivalence: VOLTA checks equivalence between optimised GPU kernels \cite{dubey2025volta}, while Gimlet explores equivalence between PyTorch specifications and generated Triton kernels \cite{taneja2026gimlet}. VeriCUDA similarly develops mechanised reasoning about Rust-to-PTX programs and reductions \cite{somani2026vericuda}.

\paragraph{Agentic verification and floating-point semantics.}
ProofWright is the closest prior work on agentic verification of generated CUDA kernels, using language models to synthesise annotations for VerCors and separately constructing semantic-equivalence proofs \cite{chatterjee2025proofwright}. In Section~\ref{sec:ArchitectureAndSetup}, we note the differences between that work and ours. Its results demonstrate that language models can effectively drive existing verification tools, but its semantic reasoning does not directly capture IEEE floating-point behaviour, instead reasoning over $\mathbb{R}$ or $\mathbb{Z}$. In addition, its LLM-generated translation from the kernel to Rocq introduces a stochastic verification surface, since the correctness of the verification target depends on an unverified and nondeterministic model-generated transformation. Recent efforts including FLoPS, FloatSpec, and fp-lean provide increasingly rich formalisations of floating-point semantics in Lean \cite{chang2026flops,floatspec2026,fplean2026}. Our work builds on this direction by checking generated GPU kernels against deterministically formalised high-level specifications while reasoning directly over floating-point values.

\begin{table}[t]
  \caption{Comparison with related GPU-kernel verification systems.
  High-level reference indicates that the system relates a kernel to an
  explicit high-level semantic reference such as a PyTorch program.
  IEEE FP indicates direct reasoning about floating-point semantics rather
  than an abstraction over $\mathbb{R}$ or $\mathbb{Z}$.}
  \label{tab:related-work}
  \centering
  \small
  \begin{tabular}{@{}lp{1.95in}cccp{1.1in}@{}}
    \toprule
    \textbf{Work} & \textbf{Brief summary} &
    \textbf{High-level ref.} &
    \textbf{IEEE FP} &
    \textbf{Agentic} &
    \textbf{Backend} \\
    \midrule

    VOLTA \citep{dubey2025volta}
    & PTX-level symbolic execution and equivalence checking for ML GPU kernels.
    & $\times$ & $\times$ & $\times$ & Custom symbolic evaluator \\

    Gimlet \citep{taneja2026gimlet}
    & PyTorch--Triton equivalence checking using tensor algebra and SMT solving.
    & $\checkmark$ & $\times$ & $\times$ & Z3 \\

    GPUVerify \citep{betts2012gpuverify}
    & Static verification of race freedom and barrier divergence in CUDA and OpenCL kernels.
    & $\times$ & $\times$ & $\times$ & Boogie / Z3 \\

    VeriCUDA \citep{somani2026vericuda}
    & Mechanised reasoning about a Rust-like GPU language and PTX-style memory-model behaviour.
    & $\times$ & $\times$ & $\times$ & Rocq \\

    VerCors \citep{safari2022vercorscuda}
    & Deductive verification of safety and functional properties of CUDA programs.
    & $\times$ & $\times$ & $\times$ & VerCors / SMT \\

    ProofWright \citep{chatterjee2025proofwright}
    & LLM-assisted verification of generated CUDA using VerCors and Rocq.
    & $\checkmark$ & $\times$ & $\checkmark$ & VerCors + Rocq \\

    \textbf{Ours}
    & Agentic semantic verification of generated CUDA against formalised high-level specifications.
    & $\checkmark$ & $\checkmark$ & $\checkmark$ & VerCors + Velvet / Lean \\

    \bottomrule
  \end{tabular}
\end{table}

\section{Architecture and Setup}
\label{sec:ArchitectureAndSetup}
\begin{figure}[ht]
    \centering
\includegraphics[width=0.85\linewidth]{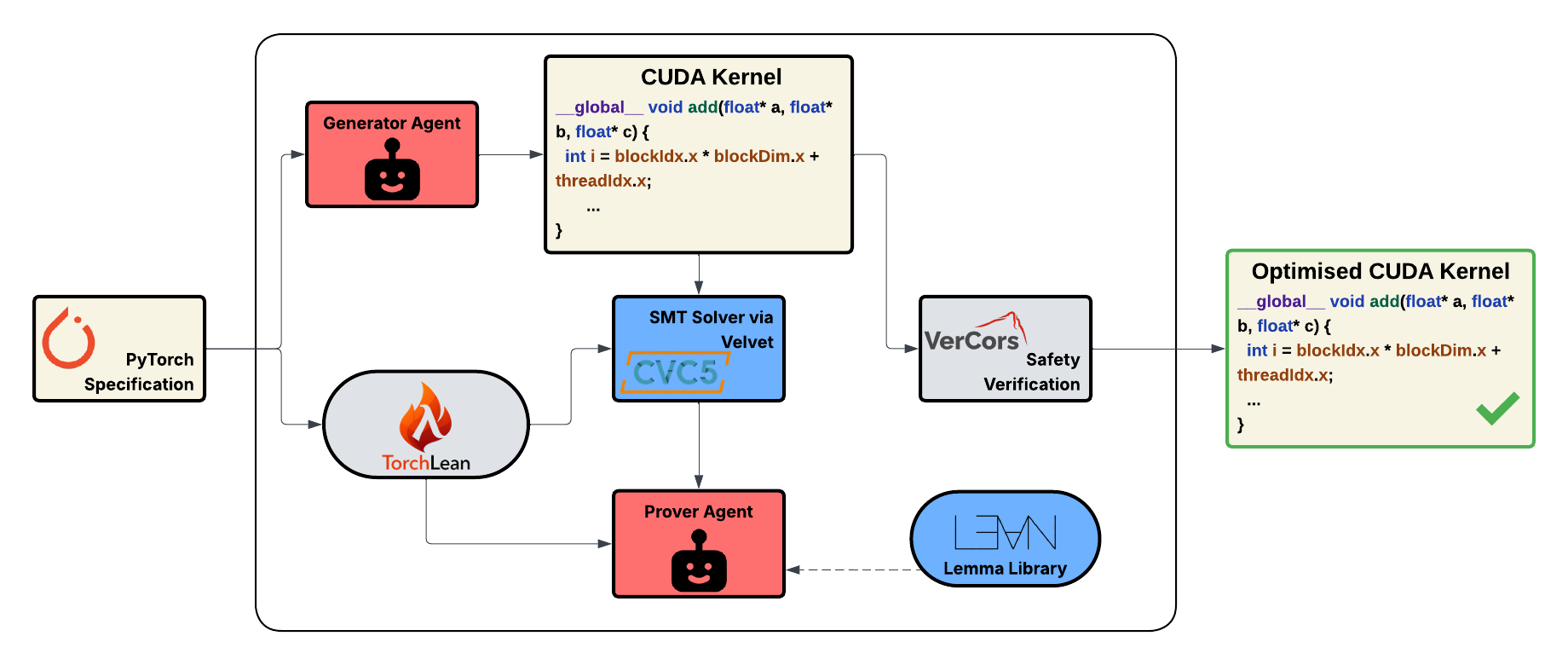}
    \caption{The basic architecture of our approach. Deterministic translators are used in order to translate both PyTorch and CUDA programs into Lean. Their semantics can then be compared either by an SMT solver (in our case, cvc5) or a Prover agent. The CUDA kernel is then checked for thread safety by VerCors.}
    \label{fig:architecture}
\end{figure}

Figure~\ref{fig:architecture} shows the basic architecture of our agentic pipeline. The setup is inspired by \cite{chatterjee2025proofwright}, with important differences reviewed at the end of this section.

\paragraph{KernelBench Problems} For a source of typical kernels, we used KernelBench \cite{ouyang2025kernelbench}, whose test-based notion of correctness and its weaknesses were discussed in Section~\ref{sec:Background}; \cite{ouyang2025kernelbench} itself names formal verification as an important next step. We should be clear that we are not aiming to make an attempt at KernelBench nor produce fast GPU kernels: the benchmark simply provides examples of kernels one would want to speed up, for which formal guarantees of semantic alignment and thread safety are valuable.

\paragraph{From PyTorch to Lean} Central to our pipeline is our ability to deterministically translate both PyTorch and CUDA kernels to semantic specifications in Lean. For the PyTorch half we employ a similar approach to \cite{chatterjee2025proofwright}, using PyTorch's \emph{torch.fx} module to extract an intermediate graph representation of the program: each applied function is a node, and edges record the order of operations and their data dependencies. This graph exposes the semantic intent of the program and can be translated deterministically into formal specifications in a theorem prover; for these specifications, we chose the TorchLean \cite{george2026torchlean} library.

\paragraph{TorchLean} Once a PyTorch program has been represented as an FX graph, its operations require a denotational semantics in Lean. TorchLean provides this semantic framework, including the \texttt{NN.IR} representation of neural-network computations through structures such as \texttt{Graph}, \texttt{Node}, and their denotation functions. We import each extracted FX graph into this representation using an explicit translation table from FX targets to TorchLean operations, with shape checking and unsupported operations reported as compilation errors. For the current KernelBench subset, we provide 30 specification modules and prove that each imported FX graph denotes the corresponding TorchLean specification, allowing the specification to be used as the trusted reference for CUDA-kernel equivalence. Thus, the comparison with a translated CUDA kernel is made against a machine-checked TorchLean semantics rather than against an unverified transcription of the PyTorch program.

\paragraph{Transcendental Functions} An obstacle for verifying facts about floating point arithmetic is the use of transcendental functions, for which IEEE recommends but does not mandate correct rounding. The main problem is the \emph{Table Maker's Dilemma} \cite{lefevre1998table}: unlike the basic operations $+, -, \times, \dots$, whose exact rounding can be decided with a small fixed number of extra bits, a transcendental's rounding can require an a priori unknown (though finite, for finite formats) amount of intermediate precision, and historically this has kept it from being mandated. In Lean v4.33.0, these functions are not checked by the kernel and are implemented as compiled code on the system's hardware. TorchLean sidesteps this verification gap by modelling tensors of an arbitrary scalar type $\alpha$: a \texttt{Context $\alpha$} instance provides the operations on $\alpha$, with the standard transcendental functions bundled in its \texttt{MathFunctions $\alpha$} component. In the case of \texttt{Float32}, we define every such function to be an \texttt{opaque} map which remains in the kernel. At first this seems restrictive in that we are unable to make use of semantic facts about the functions. For example, we cannot manipulate:
\[
    \frac{\texttt{expf(x)}}{1 + \texttt{expf(x)}} \rightarrow \frac{1}{1 + \texttt{expf(-x)}},
\]
as it relies on the relation $e^a \cdot e^b = e^{a + b}$. However, this is not an issue since in floating-point arithmetic, it does not hold for all \texttt{x} anyway. A major benefit of this approach is that we never verify anything that is false or relies on a specific hardware implementation of the transcendental functions, making the tensor representations \emph{polymorphic} over them. This guarantee should be read precisely: because both the specification and the imported kernel are written over the \emph{same} opaque symbol, equivalence is established for every interpretation of that symbol. Transferring the theorem to executed code therefore carries an additional assumption, namely that the PyTorch reference and the generated CUDA realise each such symbol as the same function on \texttt{Float32}. This is an assumption of pointwise \emph{agreement}, not of correct rounding; it holds by construction when both sides run on the same device and lower to the same library routine, but it can fail across backends, for example between a CPU reference and a GPU kernel.

\paragraph{Velvet}
Velvet \cite{gladshtein2026velvet} is a Lean library that supports the specification and verification of imperative programs. Its syntax is inspired by Dafny: a \texttt{method} may declare preconditions using \texttt{require} and postconditions using \texttt{ensures}, while its implementation is written inside a \texttt{do} block. Correctness obligations can often be discharged automatically using the \texttt{loom\_solve} tactic, which generates verification conditions and delegates suitable arithmetic reasoning to an SMT solver in a Dafny-like workflow. The result of the SMT solver is trusted in our setup. Work exists such as Lean-SMT \cite{mohamed2025lean}, for proof reconstruction of SMT output. Since Velvet is embedded in Lean, failed automated proofs can instead be completed interactively using Lean tactics and the results available in Mathlib and other imported libraries. Several of Velvet's capabilities appear to be moving into Lean itself (see, for example, the pull request \url{https://github.com/leanprover/lean4/pull/14389}), but these changes had not landed when the experiments were conducted.

\paragraph{CUDA AST Extraction}
We implemented a limited CUDA-to-Lean compiler whose frontend invokes clang to extract a CUDA abstract syntax tree and normalises it into a small kernel intermediate representation. A second stage translates this representation into a Velvet \texttt{method}, rather than into arbitrary Lean code: the generated method contains the kernel parameters, mutable arrays, sequentialised control flow, loop invariants, and proof obligations required by the Lean proving environment. The compiler assumes that the CUDA kernel is race-free and otherwise safe, and therefore replaces parallel execution with a synchronous sequential representation; barrier-separated regions are modelled as phases, while shared state and thread-local state that crosses barriers are represented explicitly. Compiler-generated preconditions primarily express operational facts such as array extents, launch dimensions, dynamic shared-memory sizes, parameter bindings, and the relationship between the thread space and the output size. The postconditions are not generally inferred from the CUDA implementation itself: they are supplied by the trusted reference specification, either as an elementwise function or as an indexed specification for reductions, convolutions, and other structured kernels. The compiler emits output-size preservation and per-element functional postconditions, together with loop invariants for bounds and already-computed output prefixes.

\paragraph{VerCors} VerCors is a deductive verification toolset for sequential, concurrent, and parallel programs, including C and GPU kernels. Programs are annotated with contracts and permission assertions; for example, a CUDA kernel may require valid input and output buffers and grant each thread fractional permissions to the locations it reads and writes. These permissions describe ownership of memory and allow VerCors to reason about pointer safety and prevent conflicting accesses between parallel threads. VerCors translates the resulting verification conditions to automated back-ends, including SMT solvers, which discharge the arithmetic, permission, and control-flow obligations. A successful verification establishes the properties expressed by the annotations, such as memory safety, data-race freedom, and, when functional postconditions are supplied, functional correctness of the modelled program. It does not by itself establish numerical equivalence to a PyTorch specification, GPU performance, or properties of CUDA behaviour that are absent from the model or annotations.

\paragraph{Generator Agent} The Generator agent is the main supervisor agent of the workflow. Its job is to take a PyTorch kernel and translate it into the corresponding CUDA code with the relevant VerCors annotations. In order to assist with this task, it is given basic advice on how to work with VerCors. Importantly, it is told to infer the \emph{weakest reasonable permissions and preconditions} that would reflect the kernel's actual accesses, since safety is conditional on these preconditions and artificially restrictive constraints would weaken the guarantee; the workflow encourages, but does not ensure, that they are the weakest possible. Upon creating a candidate CUDA kernel, the Generator agent will run the CUDA-to-Velvet compilation step creating all semantic pre- and post-conditions. Should the automatically invoked SMT solver (dispatched via \texttt{loom\_solve}) fail to prove the generated \texttt{method}, the Generator agent may dispatch the dedicated Prover agent, which is allowed to proceed with the semantic equivalence proof interactively. 

\paragraph{Prover Agent} The Prover is a specialist sub-agent of the Generator tasked with proving the equivalence of the imported PyTorch and CUDA Lean semantics. For this task, we provided it with a 300 second budget per proving round. As well as access to TorchLean, it is given two additional components upon being dispatched:
\begin{itemize}
    \item The compiled proving environment which includes the Velvet \texttt{method} and a failing invocation of the \texttt{loom\_solve} tactic,
    \item A bespoke \emph{lemma library} which it can read from and write to. These results are relevant for the problem and are accumulated across rounds.
\end{itemize}
There are two potential reasons why \texttt{loom\_solve} could fail: the method may be genuinely false, meaning the generated CUDA kernel is in fact \emph{not} equivalent to the PyTorch specification, or the SMT solver may lack the facts required to prove a true equivalence. In the first case, the agent is explicitly told to report back to the Generator with a diagnosis of the exact semantic misalignment and how the next attempt should change. In the second, it may either complete the proof interactively, continuing where \texttt{loom\_solve} left off, or add facts to the lemma library, which are made available to the SMT solver on re-compilation. The lemma library persists across rounds of a problem, but is reset upon starting a new problem so that each stays isolated and later problems do not benefit from greater automation. Both the round proof and every lemma-library addition are audited with Lean's axiom check: validation passes only if the completed proof depends on nothing beyond the three standard axioms (\texttt{propext}, \texttt{Classical.choice}, and \texttt{Quot.sound}), so the agent cannot introduce unproved assumptions.

\paragraph{Relationship to existing work}
Our approach builds on the methodology established by ProofWright \cite{chatterjee2025proofwright}, with several design differences. First, we use Lean rather than Rocq to define the denotational semantics of both PyTorch and CUDA programs, representing computational graphs with the open-source TorchLean library rather than MLRocq and giving equivalence proofs access to Mathlib \cite{TheMathlibCommunity2020TheLibrary}. Second, where ProofWright uses an LLM to translate CUDA programs into the proof environment, our CUDA-to-Lean compiler is small and deterministic: this restricts the CUDA fragment that can be represented, but makes the translation reproducible and inspectable independently of the language model. Third, verification conditions are first dispatched to the cvc5-based SMT automation \cite{BarbosaBBKLMMMN22} behind \texttt{loom\_solve}, so that straightforward equivalence proofs are discharged automatically before a Prover agent is invoked. Fourth, whereas MLRocq reportedly defines most of its tensor operations over $\mathbb{Z}$ or $\mathbb{R}$, we use TorchLean's polymorphic tensors to reason over \texttt{Float32}, so that the Lean representations match the real implementations; treating floating-point arithmetic as an implementation detail during verification is known to hide behaviour-affecting discrepancies with real-number computation \cite{goldberg1991every, monniaux2008pitfalls}. Finally, rather than maintaining a dedicated VerCors knowledge base as \cite{chatterjee2025proofwright} do, we include only minimal VerCors advice in the Generator agent's prompt, speculating that newer underlying models have likely been trained on the relevant content.


\section{Results}
\label{sec:Results}
\begin{figure}[t]
    \centering

    \begin{subfigure}{0.48\linewidth}
        \centering
        \includegraphics[width=\linewidth]{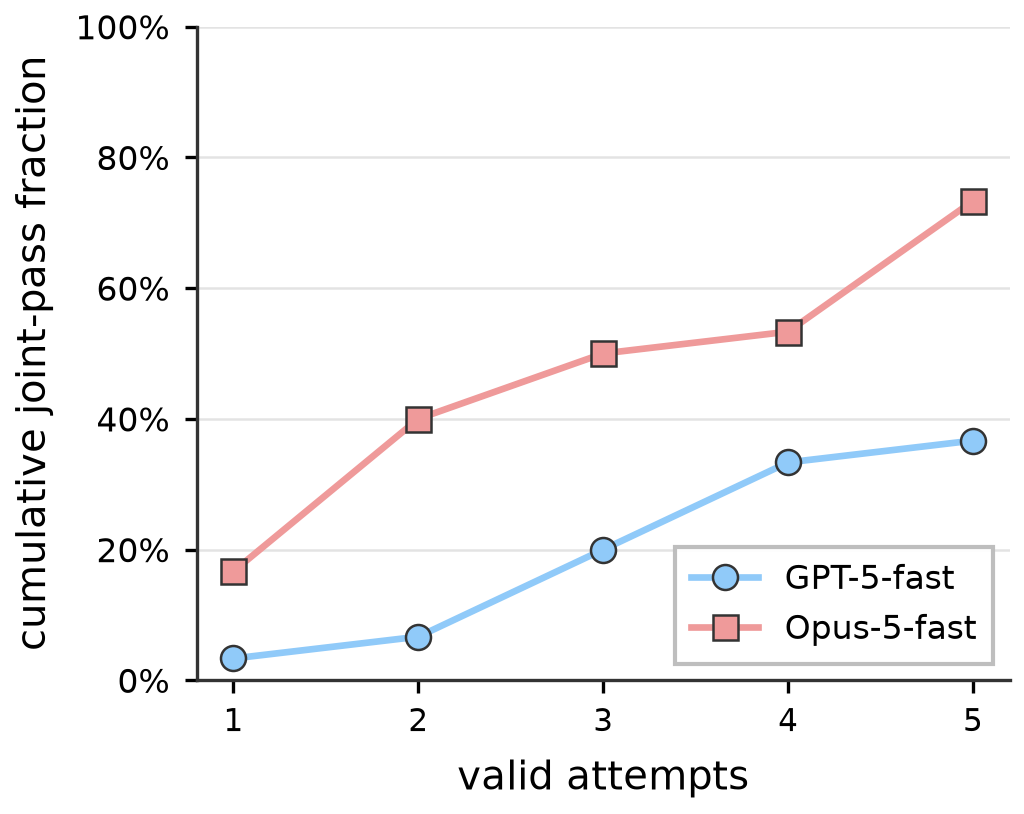}
        \caption{Cumulative joint pass rate}
        \label{fig:conversion}
    \end{subfigure}
    \hfill
    \begin{subfigure}{0.48\linewidth}
        \centering
        \includegraphics[width=\linewidth]{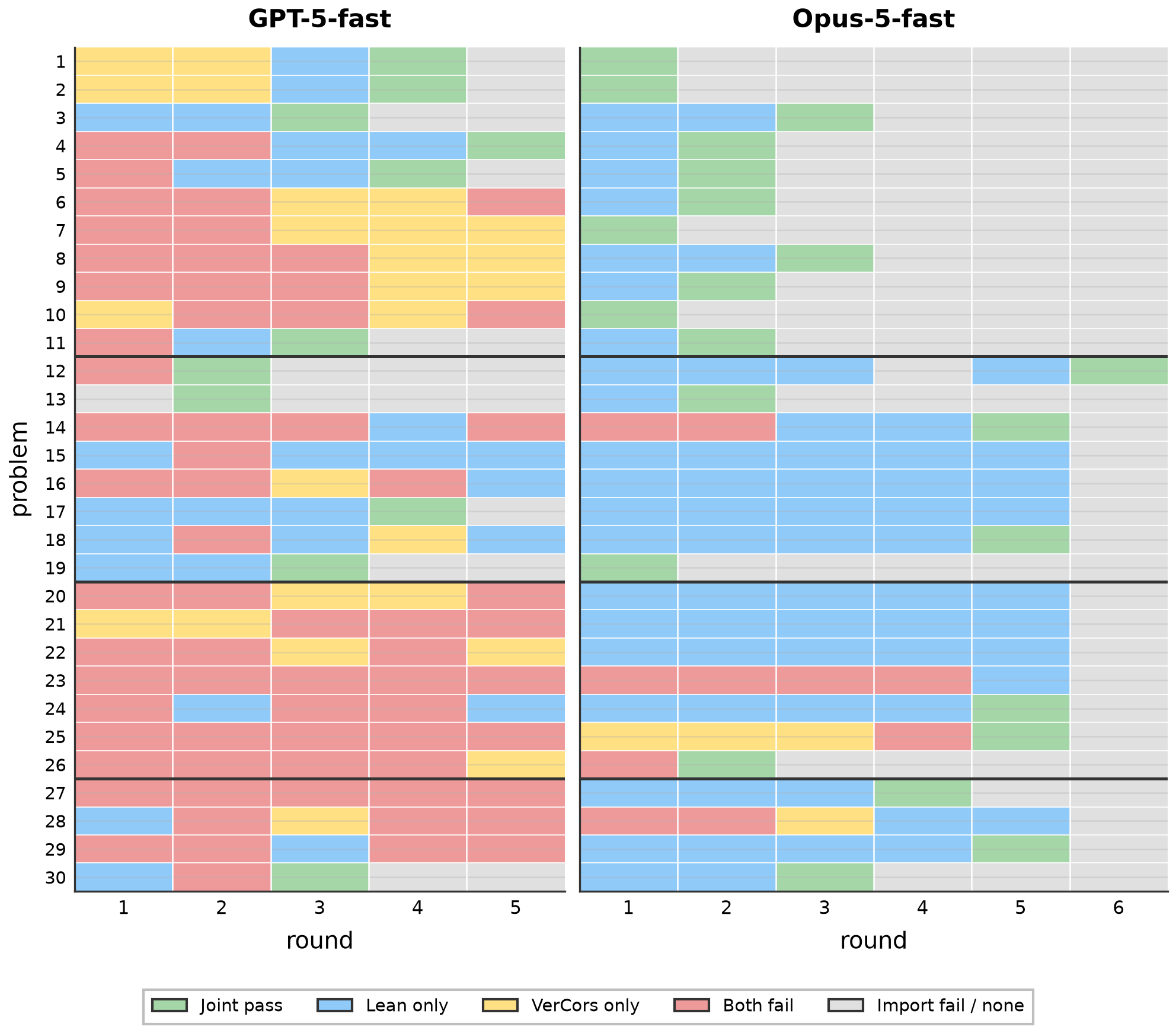}
        \caption{Per-round outcome grid}
        \label{fig:outcome-grid}
    \end{subfigure}

    \caption{Pass rates for each model across all problems. Figure~\ref{fig:conversion} shows the cumulative fraction of problems jointly verified against the number of valid attempts used; Opus-5-fast ultimately validates roughly double the number of kernels that GPT-5-fast does. Figure~\ref{fig:outcome-grid} shows, for every problem (indexed $1$--$30$ down the left, grouped by family) and every round, the round's outcome: both axes passed, only the semantic (Lean) axis, only the safety (VerCors) axis, neither, or an invalid round whose import failed.}
    \label{fig:combined}
\end{figure}

\paragraph{The Experiment} We ran our pipeline with different LLMs taking the place of the Generator and Prover agent. Each undertook the 30 selected problems from levels 1 and 2 of KernelBench. These are problems whose reference computations lie within the fragment supported by our deterministic compilers (Section~\ref{sec:ArchitectureAndSetup}); within that fragment we chose problems to be representative, spanning the four structural families of Table~\ref{tab:aggregate} (the full list is given in Appendix~\ref{app:problems}), and 30 is enough to give a meaningful picture of the pipeline's capabilities. For a given LLM and a given problem, the agent is allowed up to 5 valid \emph{rounds} in order to produce a semantically correct and memory safe kernel. A round constitutes the following steps. The Generator agent is provided with the original PyTorch code and asked to generate the corresponding CUDA code with VerCors annotations. If importing this into the proving environment via the CUDA AST Extraction process is successful, the round counts as \emph{valid}. In order that the LLM is not restricted by limitations of the compiler, an \emph{invalid} round is allowed to be retried. Upon this import being successful, the proving phase begins, invoking the \texttt{loom\_solve} SMT tactic and an optional deployment of the Prover agent. Regardless of the outcome of the proving phase, VerCors proceeds to test the kernel for safety. If both of these phases are successful, the round and problem are complete. If either fails, all feedback is given back to the Generator agent who begins a fresh round for the problem. All successes, failures, timings and intermediate kernels were logged throughout the process. We made use of OpenCode \cite{opencode} in order to implement the workflow. The pipeline requires no GPU: every verification stage (the CUDA import, Lean elaboration, cvc5, and VerCors) runs on CPU on a single workstation, with the LLMs accessed through their public APIs; the campaign totalled roughly 14 hours of pipeline wall-clock time over 131 rounds for GPT-5-fast and 9 hours over 102 rounds for Opus-5-fast. Figure~\ref{fig:combined} shows the overall pass rates for both models over each round.

 

\begin{table}[t]
  \centering
  \caption{Joint (Lean~$\wedge$~VerCors) and per-axis pass rates within five valid attempts, by specification family. Counts are passed/total problems; percentages in parentheses. Notably, Claude Opus 5 is able to provide semantically equivalent kernels for every problem, but neither model is able to prove thread safety for all problems. A Lean proof here is a semantic equivalence between the adapted, sequential version of the kernel, not the original CUDA implementation. We are trusting that this translation is accurate. }
  \label{tab:aggregate}
  \small
  \begin{tabular}{l r r r r r r r}
    \toprule
    & & \multicolumn{3}{c}{GPT-5-fast} & \multicolumn{3}{c}{Opus-5-fast} \\
    \cmidrule(lr){3-5} \cmidrule(lr){6-8}
    Family & $n$ & Joint & Lean & VerCors & Joint & Lean & VerCors \\
    \midrule
    Elementwise & 11 & 6 (55\%) & 6 (55\%) & 11 (100\%) & 11 (100\%) & 11 (100\%) & 11 (100\%) \\
    Matmul & 8 & 4 (50\%) & 8 (100\%) & 6 (75\%) & 5 (62\%) & 8 (100\%) & 5 (62\%) \\
    Gemm+epilogue & 7 & 0 (0\%) & 1 (14\%) & 4 (57\%) & 3 (43\%) & 7 (100\%) & 3 (43\%) \\
    Reduction & 4 & 1 (25\%) & 3 (75\%) & 2 (50\%) & 3 (75\%) & 4 (100\%) & 4 (100\%) \\
    \midrule
    \textbf{Overall} & 30 & 11 (37\%) & 18 (60\%) & 23 (77\%) & 22 (73\%) & 30 (100\%) & 23 (77\%) \\
    \bottomrule
  \end{tabular}
\end{table}

\paragraph{Problem categories} We group the 30 problems into four families according to the structure of the reference computation, since this structure, rather than the surface identity of the operator, is what determines the shape of the resulting proof. \emph{Elementwise} problems (the activation functions) compute each output from the single corresponding input, so the contract is a per-element scalar equality with no cross-element dependence. \emph{Matmul} problems compute each output as a dot product over a shared dimension, introducing an accumulation loop whose partial sums must be captured by a loop invariant and whose summation order is part of the specification. \emph{Gemm+epilogue} problems (the level-2 kernels) compose a matrix multiply with a bias and an elementwise epilogue, so their proofs combine the accumulation reasoning of a matmul with the exact-form reasoning of an activation. \emph{Reduction} problems (softmax and the norms) make each output depend on an entire row or on the whole tensor, requiring one or more reduction passes before any output can be written. Table~\ref{tab:aggregate} reports pass rates within these families and Figure~\ref{fig:outcome-grid} resolves them to the level of individual rounds.

\paragraph{Analysis} Two patterns are visible across Table~\ref{tab:aggregate} and Figure~\ref{fig:outcome-grid}. First, the two axes fail for different reasons and in different places. The semantic axis is where the models differ most: Opus-5-fast establishes semantic equivalence on all thirty problems, whereas GPT-5-fast does so on eighteen, and essentially all of that gap sits in the structured families. On the elementwise problems the difference is one of exact floating-point form, GPT-5-fast frequently producing a kernel that is close but not bit-identical to the reference (Section~\ref{sec:cs-semantic}), which costs it five of eleven problems on the semantic axis while Opus-5-fast loses none. The gap widens with structural complexity: on the gemm+epilogue family, which demands both accumulation and epilogue reasoning at once, GPT-5-fast establishes equivalence on only one of seven problems against a clean sweep for Opus-5-fast. The safety axis, by contrast, is largely model-independent: both models verify safety on the same twenty-three problems, and it is the binding constraint on the matmul and gemm families, where the structured access patterns make the required memory-safety annotations hardest to get right regardless of which model produced them. Second, the joint metric is therefore gated by whichever axis is harder for a given family: elementwise and reduction problems are limited by the semantic axis and reduction problems by it only for the weaker model, whereas matmul and gemm are limited by safety. The net effect is that Opus-5-fast, by clearing the semantic axis everywhere, converts most of its safety passes into joint passes, while GPT-5-fast forfeits joint passes it could otherwise have claimed on the structured families because its kernels are not semantically faithful. We caution that these observations derive from a single campaign per model under one agent configuration; they characterise our runs rather than constituting a controlled comparison of the underlying models.




\FloatBarrier  

\section{Case Study}
\label{sec:CaseStudy}

We now examine two representative case studies from the runs. The first shows the semantic checker certifying a kernel whose numerics depend on a transcendental function that the framework never interprets. The second shows the checker rejecting a kernel over a genuine floating-point discrepancy, and the feedback loop repairing the CUDA implementation. Three further studies are deferred to Appendix~\ref{app:casestudies}. Throughout, we refer to problems by their index (1--30) in the ordering of Figure~\ref{fig:outcome-grid}; the indexed problem list is given in Appendix~\ref{app:problems}.

\subsection{Verifying floating-point numerics through opaque functions}
\label{sec:cs-opaque}

On the sigmoid activation (problem~3), both GPT-5-fast and Opus-5-fast converged quickly, each writing essentially the kernel below and having its equivalence to the PyTorch reference discharged automatically by \texttt{loom\_solve}, with no interactive proof and no additions to the lemma library.
\begin{SaveVerbatim}{originalcode1}
__global__ void sigmoid_kernel(float* x, float* out, int n) {
  int i = blockIdx.x * blockDim.x + threadIdx.x;
  if (i < n) out[i] = 1.0f / (1.0f + expf(-x[i]));   // spec: 1 / (1 + exp(-x))
}
\end{SaveVerbatim}
\begin{Verbatim}[numbersep=5pt,frame=lines,framesep=2mm,commandchars=\\\{\}]
\PYG{k+kr}{\PYGZus{}\PYGZus{}global\PYGZus{}\PYGZus{}}\PYG{+w}{ }\PYG{k+kt}{void}\PYG{+w}{ }\PYG{n}{sigmoid\PYGZus{}kernel}\PYG{p}{(}\PYG{k+kt}{float}\PYG{o}{*}\PYG{+w}{ }\PYG{n}{x}\PYG{p}{,}\PYG{+w}{ }\PYG{k+kt}{float}\PYG{o}{*}\PYG{+w}{ }\PYG{n}{out}\PYG{p}{,}\PYG{+w}{ }\PYG{k+kt}{int}\PYG{+w}{ }\PYG{n}{n}\PYG{p}{)}\PYG{+w}{ }\PYG{p}{\PYGZob{}}
\PYG{+w}{  }\PYG{k+kt}{int}\PYG{+w}{ }\PYG{n}{i}\PYG{+w}{ }\PYG{o}{=}\PYG{+w}{ }\PYG{n+nb}{blockIdx}\PYG{p}{.}\PYG{n}{x}\PYG{+w}{ }\PYG{o}{*}\PYG{+w}{ }\PYG{n+nb}{blockDim}\PYG{p}{.}\PYG{n}{x}\PYG{+w}{ }\PYG{o}{+}\PYG{+w}{ }\PYG{n+nb}{threadIdx}\PYG{p}{.}\PYG{n}{x}\PYG{p}{;}
\PYG{+w}{  }\PYG{k}{if}\PYG{+w}{ }\PYG{p}{(}\PYG{n}{i}\PYG{+w}{ }\PYG{o}{\PYGZlt{}}\PYG{+w}{ }\PYG{n}{n}\PYG{p}{)}\PYG{+w}{ }\PYG{n}{out}\PYG{p}{[}\PYG{n}{i}\PYG{p}{]}\PYG{+w}{ }\PYG{o}{=}\PYG{+w}{ }\PYG{l+m+mf}{1.0f}\PYG{+w}{ }\PYG{o}{/}\PYG{+w}{ }\PYG{p}{(}\PYG{l+m+mf}{1.0f}\PYG{+w}{ }\PYG{o}{+}\PYG{+w}{ }\PYG{n}{expf}\PYG{p}{(}\PYG{o}{\PYGZhy{}}\PYG{n}{x}\PYG{p}{[}\PYG{n}{i}\PYG{p}{]));}\PYG{+w}{   }\PYG{c+c1}{// spec: 1 / (1 + exp(\PYGZhy{}x))}
\PYG{p}{\PYGZcb{}}
\end{Verbatim}

The kernel calls \texttt{expf}, and CUDA's device \texttt{expf} has no bit-exact model in Lean. As described in Section~\ref{sec:ArchitectureAndSetup}, the compiler therefore maps it to an \emph{opaque} symbol, and the reference specification is written over the \emph{same} symbol. The proof does not compute an exponential anywhere: it succeeds because the kernel applies that symbol to the same argument ($-x_i$) as the specification, and the remaining operations, the addition and the reciprocal, are the fully modelled \texttt{Float32} operations, which the two sides perform identically. What is established is thus a structural equivalence that holds \emph{for every interpretation} of the exponential: the pipeline never asserts anything about how the hardware rounds \texttt{expf}, and never verifies a numerical identity that is false in floating point.

The one instructive divergence concerned \emph{which} function is called. On the ELU activation (problem~9), GPT-5-fast substituted the fast-math intrinsic \texttt{\_\_expf} for \texttt{expf}; because the approximate intrinsic is imported as a \emph{distinct} opaque symbol, the equivalence could not be established, and the model reverted to \texttt{expf} in the following round. This is a genuine mistake surfaced by the checker, since \texttt{\_\_expf} is not interchangeable with \texttt{expf} at the bit level.

\subsection{Feedback prompting a semantic correction}
\label{sec:cs-semantic}

Interesting cases are those in which the semantic-equivalence step did not pass and the feedback identified the cause as a genuine discrepancy rather than an inability of the Prover agent to complete the proof. A subtle example arose for Opus-5-fast on problem~25, a fused matrix-multiply with a bias and additive epilogue followed by a ReLU. For its first three rounds the kernel wrote the rectification as a strict test against zero,
\begin{SaveVerbatim}{originalcode2}
out[i] = (y < 0.0f) ? 0.0f : y;                   // rounds 1-3: equivalence did not pass
out[i] = (y > 0.0f) ? y : ((y != y) ? y : 0.0f);  // round 5: accepted
\end{SaveVerbatim}
\begin{Verbatim}[numbersep=5pt,frame=lines,framesep=2mm,commandchars=\\\{\}]
\PYG{n}{out}\PYG{p}{[}\PYG{n}{i}\PYG{p}{]}\PYG{+w}{ }\PYG{o}{=}\PYG{+w}{ }\PYG{p}{(}\PYG{n}{y}\PYG{+w}{ }\PYG{o}{\PYGZlt{}}\PYG{+w}{ }\PYG{l+m+mf}{0.0f}\PYG{p}{)}\PYG{+w}{ }\PYG{o}{?}\PYG{+w}{ }\PYG{l+m+mf}{0.0f}\PYG{+w}{ }\PYG{o}{:}\PYG{+w}{ }\PYG{n}{y}\PYG{p}{;}\PYG{+w}{                   }\PYG{c+c1}{// rounds 1\PYGZhy{}3: equivalence did not pass}
\PYG{n}{out}\PYG{p}{[}\PYG{n}{i}\PYG{p}{]}\PYG{+w}{ }\PYG{o}{=}\PYG{+w}{ }\PYG{p}{(}\PYG{n}{y}\PYG{+w}{ }\PYG{o}{\PYGZgt{}}\PYG{+w}{ }\PYG{l+m+mf}{0.0f}\PYG{p}{)}\PYG{+w}{ }\PYG{o}{?}\PYG{+w}{ }\PYG{n}{y}\PYG{+w}{ }\PYG{o}{:}\PYG{+w}{ }\PYG{p}{((}\PYG{n}{y}\PYG{+w}{ }\PYG{o}{!=}\PYG{+w}{ }\PYG{n}{y}\PYG{p}{)}\PYG{+w}{ }\PYG{o}{?}\PYG{+w}{ }\PYG{n}{y}\PYG{+w}{ }\PYG{o}{:}\PYG{+w}{ }\PYG{l+m+mf}{0.0f}\PYG{p}{);}\PYG{+w}{  }\PYG{c+c1}{// round 5: accepted}
\end{Verbatim}

Over the reals the strict form is exactly the rectifier, but in \texttt{Float32} it disagrees with the reference on a single input: applied to negative zero, the test \texttt{-0.0f < 0.0f} is false, so the kernel returns \texttt{-0.0f}, whereas the reference's ReLU maps \texttt{-0.0} to \texttt{+0.0}. This meant the checker declined the kernel. The accepted form passes strictly positive values through unchanged, propagates \texttt{NaN} via the \texttt{y != y} test, and sends negatives and both zeros to \texttt{+0.0f}. The model's own kernel comments record this diagnosis, noting also that \texttt{fmaxf} is deliberately avoided because IEEE's maximum-number operation would drop the \texttt{NaN}s the specification propagates (cf.\ Appendix~\ref{app:cs-unresolved}). An intermediate round wrote the non-strict \texttt{(y <= 0.0f) ? 0.0f : y}, which no longer disagreed with the reference on any input but was still rejected: it is not a form the automation could bridge to the specification's maximum, a provability failure of the kind discussed next. Because both models write real \texttt{Float32} arithmetic that the framework models, the pipeline is able to distinguish behaviour on special values that finite tests are unlikely to pick up on.


\FloatBarrier  

\section{Conclusion}
\label{sec:Conclusion}

In this work, we introduced a framework for verifying LLM-generated GPU kernels from a high-level PyTorch specification. The framework employs various technologies such as Velvet, VerCors and TorchLean in order to provide semantic and safety guarantees about the LLM-generated parallel CUDA code. Our work aimed to build upon existing work such as \cite{chatterjee2025proofwright} by introducing two key differences:
\begin{enumerate}
    \item \label{enum:1} A minimal, deterministic CUDA-to-Lean translation step, rather than an LLM, allowing for semantic equivalence to be ascertained between the CUDA and PyTorch code.
    \item \label{enum:2} Semantic verifications were performed over \texttt{Float32}, with transcendental functions implemented as \texttt{opaque} definitions.
\end{enumerate}
The effect of \ref{enum:1} is that we reduce the amount of CUDA code we can reason about in Lean. However, it means that if an incorrect CUDA-to-Lean translation takes place, it is only the compiler that needs to be inspected. It can be modified and fixed. The importance of \ref{enum:2} is that treating numerical algorithms as if they were implemented over $\mathbb{R}$ or $\mathbb{Z}$ can lead to large errors, thus rendering the semantic verification less meaningful. We provided case studies in Section~\ref{sec:CaseStudy} and Appendix~\ref{app:casestudies} that demonstrate concrete examples of how our harness can catch these semantic discrepancies. 

\paragraph{Limitations} One limitation of our approach is that whilst a deterministic CUDA-to-Lean transformation step is valuable for removing stochasticity from the workflow, it means that the semantics that we can represent are limited by that compiler. Rather than relying on this or an LLM, another approach could be similar to that used in \cite{hietala2024lean}: providing an executable semantics in Lean for the target program, then relying on differential testing to ensure validity of the proposed semantics. A second limitation is that as it currently stands, there are no semantics defined for hardware-level transcendental functions in Lean. Because these functions are left opaque, our equivalence theorems transfer to executed code only under the assumption that the specification-side and kernel-side implementations of each transcendental agree pointwise (Section~\ref{sec:ArchitectureAndSetup}); when the reference and the kernel run on the same device this holds by construction, but across backends the two implementations need not agree bit-for-bit, and the guarantee is then equivalence only up to a common interpretation of those functions. Giving these functions genuine semantics in Lean would be a difficult challenge but an important next step if we are to trust the results of LLM-generated and optimised GPU code. Finally, our evaluation is a single campaign per model over 30 problems: pass rates are reported as raw counts, and we make no attempt to characterise run-to-run variance in agent behaviour.

\paragraph{Broader impact} Machine-checked verification of LLM-generated kernels offers a route to trusting generated code in production ML systems: it catches silent numerical defects that empirical testing misses (Section~\ref{sec:CaseStudy}), and as code generation is increasingly delegated to language models, we regard establishing this kind of trust as the principal positive impact of this line of work. The corresponding risk is over-trust: an equivalence certificate covers only the stated specification and assumptions, and could lend unearned confidence to code whose preconditions are violated in deployment or whose transcendental implementations differ between the specification and target backends. We mitigate this by keeping every assumption explicit in the proving environment and by stating precisely the conditions under which our guarantees transfer to executed code (Section~\ref{sec:ArchitectureAndSetup}).

\makeatletter
\if@anonymous\else
\begin{ack}
This work was carried out at Logos Research. We are grateful to Logos Research for the compute resources, engineering infrastructure, and research environment that supported this work and made the work possible, and to our colleagues there for helpful discussions and feedback.
\end{ack}
\fi
\makeatother

\bibliographystyle{plainnat}
\bibliography{references}
\appendix

\section{The 30 benchmark problems}
\label{app:problems}

Table~\ref{tab:problems} lists the 30 KernelBench problems used in the campaign, in the index order of Figure~\ref{fig:outcome-grid}: grouped by structural family (Section~\ref{sec:Results}), then by KernelBench problem number.

\begin{table}[ht]
  \caption{The 30 KernelBench problems attempted by both models, with their KernelBench level and structural family.}
  \label{tab:problems}
  \centering
  \small
  \begin{tabular}{@{}r c l l@{}}
    \toprule
    \# & Level & KernelBench problem & Family \\
    \midrule
    1  & 1 & 19\_ReLU & Elementwise \\
    2  & 1 & 20\_LeakyReLU & Elementwise \\
    3  & 1 & 21\_Sigmoid & Elementwise \\
    4  & 1 & 22\_Tanh & Elementwise \\
    5  & 1 & 25\_Swish & Elementwise \\
    6  & 1 & 27\_SELU & Elementwise \\
    7  & 1 & 28\_HardSigmoid & Elementwise \\
    8  & 1 & 30\_Softsign & Elementwise \\
    9  & 1 & 31\_ELU & Elementwise \\
    10 & 1 & 32\_HardTanh & Elementwise \\
    11 & 1 & 88\_MinGPTNewGelu & Elementwise \\
    12 & 1 & 1\_Square\_matrix\_multiplication & Matmul \\
    13 & 1 & 2\_Standard\_matrix\_multiplication & Matmul \\
    14 & 1 & 4\_Matrix\_vector\_multiplication & Matmul \\
    15 & 1 & 6\_Matmul\_with\_large\_K\_dimension & Matmul \\
    16 & 1 & 7\_Matmul\_with\_small\_K\_dimension & Matmul \\
    17 & 1 & 8\_Matmul\_with\_irregular\_shapes & Matmul \\
    18 & 1 & 9\_Tall\_skinny\_matrix\_multiplication & Matmul \\
    19 & 1 & 13\_Matmul\_for\_symmetric\_matrices & Matmul \\
    20 & 2 & 9\_Matmul\_Subtract\_Multiply\_ReLU & Gemm+epilogue \\
    21 & 2 & 12\_Gemm\_Multiply\_LeakyReLU & Gemm+epilogue \\
    22 & 2 & 59\_Matmul\_Swish\_Scaling & Gemm+epilogue \\
    23 & 2 & 63\_Gemm\_ReLU\_Divide & Gemm+epilogue \\
    24 & 2 & 70\_Gemm\_Sigmoid\_Scaling\_ResidualAdd & Gemm+epilogue \\
    25 & 2 & 76\_Gemm\_Add\_ReLU & Gemm+epilogue \\
    26 & 2 & 81\_Gemm\_Swish\_Divide\_Clamp\_Tanh\_Clamp & Gemm+epilogue \\
    27 & 1 & 23\_Softmax & Reduction \\
    28 & 1 & 37\_FrobeniusNorm & Reduction \\
    29 & 1 & 38\_L1Norm & Reduction \\
    30 & 1 & 39\_L2Norm & Reduction \\
    \bottomrule
  \end{tabular}
\end{table}

\section{Additional case studies}
\label{app:casestudies}

This appendix collects three further episodes referenced in Section~\ref{sec:CaseStudy}: a safety-driven repair, a semantic rejection that the model did not resolve, and a proof obligation that recurred across almost every problem.

\subsection{Feedback prompting a safety correction}
\label{app:cs-safety}

Whereas Section~\ref{sec:cs-semantic} concerns the semantic axis, the safety axis produced its own repairs. On problem~13, a standard matrix multiplication, Opus-5-fast's first kernel was already memory-safe. Each thread wrote a single distinct output cell, but the safety checker could not confirm it, reporting that a read offset might fall outside the buffer. The kernel body was left unchanged in the next round; the model instead supplied the buffer extents it had previously omitted and restated them as loop invariants:
\begin{SaveVerbatim}{originalcode3}
/*@ context \pointer_length(A) >= 16777216;      // added: buffer extents
    context \pointer_length(B) >= 33554432;
    context \pointer_length(C) >= 8388608;   ... @*/
for (int k = 0; k < K; k++) {
  /*@ loop_invariant \pointer_length(A) >= 16777216; ... @*/   // carried into loop
  acc += A[r*K + k] * B[k*N + c];
}
\end{SaveVerbatim}
\begin{Verbatim}[numbersep=5pt,frame=lines,framesep=2mm,commandchars=\\\{\}]
\PYG{c+cm}{/*@ context \PYGZbs{}pointer\PYGZus{}length(A) \PYGZgt{}= 16777216;      // added: buffer extents}
\PYG{c+cm}{    context \PYGZbs{}pointer\PYGZus{}length(B) \PYGZgt{}= 33554432;}
\PYG{c+cm}{    context \PYGZbs{}pointer\PYGZus{}length(C) \PYGZgt{}= 8388608;   ... @*/}
\PYG{k}{for}\PYG{+w}{ }\PYG{p}{(}\PYG{k+kt}{int}\PYG{+w}{ }\PYG{n}{k}\PYG{+w}{ }\PYG{o}{=}\PYG{+w}{ }\PYG{l+m+mi}{0}\PYG{p}{;}\PYG{+w}{ }\PYG{n}{k}\PYG{+w}{ }\PYG{o}{\PYGZlt{}}\PYG{+w}{ }\PYG{n}{K}\PYG{p}{;}\PYG{+w}{ }\PYG{n}{k}\PYG{o}{++}\PYG{p}{)}\PYG{+w}{ }\PYG{p}{\PYGZob{}}
\PYG{+w}{  }\PYG{c+cm}{/*@ loop\PYGZus{}invariant \PYGZbs{}pointer\PYGZus{}length(A) \PYGZgt{}= 16777216; ... @*/}\PYG{+w}{   }\PYG{c+c1}{// carried into loop}
\PYG{+w}{  }\PYG{n}{acc}\PYG{+w}{ }\PYG{o}{+=}\PYG{+w}{ }\PYG{n}{A}\PYG{p}{[}\PYG{n}{r}\PYG{o}{*}\PYG{n}{K}\PYG{+w}{ }\PYG{o}{+}\PYG{+w}{ }\PYG{n}{k}\PYG{p}{]}\PYG{+w}{ }\PYG{o}{*}\PYG{+w}{ }\PYG{n}{B}\PYG{p}{[}\PYG{n}{k}\PYG{o}{*}\PYG{n}{N}\PYG{+w}{ }\PYG{o}{+}\PYG{+w}{ }\PYG{n}{c}\PYG{p}{];}
\PYG{p}{\PYGZcb{}}
\end{Verbatim}

The accesses were in bounds throughout; the difficulty was that the reasoning could not \emph{derive} that fact without the extents being asserted and, because the checker does not propagate facts across a loop automatically, restated as invariants. The feedback therefore taught the model to supply a true precondition the verifier required rather than to alter the computation. This is a completeness gap in the annotations, made visible and closed by the loop, rather than a defect in the kernel; annotation-only safety repairs of this kind were common across both models.

\subsection{A semantic mismatch that was not resolved}
\label{app:cs-unresolved}

Not every rejection was repaired, and the row-wise softmax (problem~27) provides an informative comparison between the two models. The specification computes a numerically stable softmax, subtracting the row maximum before exponentiating; the maximum is a fold using the reference's maximum operation, which propagates \texttt{NaN}. GPT-5-fast implemented the fold with the CUDA library call \texttt{fmaxf}:
\begin{SaveVerbatim}{originalcode4}
rowmax = fmaxf(rowmax, x[base + k]);        // GPT-5-fast: equivalence never passed
rowmax = (rowmax <= x[base + k]) ? x[base + k] : rowmax;   // Opus-5-fast: accepted
\end{SaveVerbatim}
\begin{Verbatim}[numbersep=5pt,frame=lines,framesep=2mm,commandchars=\\\{\}]
\PYG{n}{rowmax}\PYG{+w}{ }\PYG{o}{=}\PYG{+w}{ }\PYG{n}{fmaxf}\PYG{p}{(}\PYG{n}{rowmax}\PYG{p}{,}\PYG{+w}{ }\PYG{n}{x}\PYG{p}{[}\PYG{n}{base}\PYG{+w}{ }\PYG{o}{+}\PYG{+w}{ }\PYG{n}{k}\PYG{p}{]);}\PYG{+w}{        }\PYG{c+c1}{// GPT\PYGZhy{}5\PYGZhy{}fast: equivalence never passed}
\PYG{n}{rowmax}\PYG{+w}{ }\PYG{o}{=}\PYG{+w}{ }\PYG{p}{(}\PYG{n}{rowmax}\PYG{+w}{ }\PYG{o}{\PYGZlt{}=}\PYG{+w}{ }\PYG{n}{x}\PYG{p}{[}\PYG{n}{base}\PYG{+w}{ }\PYG{o}{+}\PYG{+w}{ }\PYG{n}{k}\PYG{p}{])}\PYG{+w}{ }\PYG{o}{?}\PYG{+w}{ }\PYG{n}{x}\PYG{p}{[}\PYG{n}{base}\PYG{+w}{ }\PYG{o}{+}\PYG{+w}{ }\PYG{n}{k}\PYG{p}{]}\PYG{+w}{ }\PYG{o}{:}\PYG{+w}{ }\PYG{n}{rowmax}\PYG{p}{;}\PYG{+w}{   }\PYG{c+c1}{// Opus\PYGZhy{}5\PYGZhy{}fast: accepted}
\end{Verbatim}

This is a genuine bit-level discrepancy and it is surfaced by the checker: \texttt{fmaxf} realises the IEEE ``maximum-number'' operation, which \emph{drops} a \texttt{NaN} operand, whereas the reference's maximum propagates it, so the two disagree on any row containing a \texttt{NaN}. Because the contract quantifies over all inputs, the semantic-equivalence step correctly declined the kernel, and it did so in all five of GPT-5-fast's attempts. Opus-5-fast, by contrast, wrote the reference's exact ternary maximum from its first attempt. Its kernel comments explicitly note that this differs from \texttt{fmaxf}, and its softmax passed the semantic step in every round. This case illustrates that the checker distinguishes two functions that agree on all non-\texttt{NaN} inputs.

\subsection{Recurring proof obligations and the case for a standing library}
\label{app:cs-library}

A striking feature of the runs is how often the Prover agent re-established the same facts in the lemma library. Because the lemma library is reset at the start of each problem (Section~\ref{sec:ArchitectureAndSetup}), any general-purpose fact must be re-proved from scratch for every problem that needs it. One obligation in particular dominated: a book-keeping fact about the output array, namely that extending a per-cell postcondition across a single write preserves it, so that if every already-written cell matches the specification then it still does after one further cell is written. In the runs this was stated per problem in the form below (shown for the rectifier of problem~1).
\begin{SaveVerbatim}{originalcode5}
theorem set_extend (x out : Array Float32) (gid : Nat) (v : Float32)
    (hlt : gid < out.size)
    (hv : v = spec x[gid]!)
    (hprev : $\forall$ k, k < gid $\to$ out[k]! = spec x[k]!) :
    $\forall$ k, k < gid + 1 $\to$ (out.set! gid v)[k]! = spec x[k]!
\end{SaveVerbatim}
\begin{Verbatim}[numbersep=5pt,frame=lines,framesep=2mm,commandchars=\\\{\},codes={\catcode`\$=3\catcode`\^=7\catcode`\_=8\relax}]
\PYG{k+kd}{theorem}\PYG{+w}{ }\PYG{n}{set\PYGZus{}extend}\PYG{+w}{ }\PYG{o}{(}\PYG{n}{x}\PYG{+w}{ }\PYG{n}{out}\PYG{+w}{ }\PYG{o}{:}\PYG{+w}{ }\PYG{n}{Array}\PYG{+w}{ }\PYG{n}{Float32}\PYG{o}{)}\PYG{+w}{ }\PYG{o}{(}\PYG{n}{gid}\PYG{+w}{ }\PYG{o}{:}\PYG{+w}{ }\PYG{n}{Nat}\PYG{o}{)}\PYG{+w}{ }\PYG{o}{(}\PYG{n}{v}\PYG{+w}{ }\PYG{o}{:}\PYG{+w}{ }\PYG{n}{Float32}\PYG{o}{)}
\PYG{+w}{    }\PYG{o}{(}\PYG{n}{hlt}\PYG{+w}{ }\PYG{o}{:}\PYG{+w}{ }\PYG{n}{gid}\PYG{+w}{ }\PYG{n+nb+bp}{\PYGZlt{}}\PYG{+w}{ }\PYG{n}{out.size}\PYG{o}{)}
\PYG{+w}{    }\PYG{o}{(}\PYG{n}{hv}\PYG{+w}{ }\PYG{o}{:}\PYG{+w}{ }\PYG{n}{v}\PYG{+w}{ }\PYG{n+nb+bp}{=}\PYG{+w}{ }\PYG{n}{spec}\PYG{+w}{ }\PYG{n}{x}\PYG{o}{[}\PYG{n}{gid}\PYG{o}{]}\PYG{n+nb+bp}{!}\PYG{o}{)}
\PYG{+w}{    }\PYG{o}{(}\PYG{n}{hprev}\PYG{+w}{ }\PYG{o}{:}\PYG{+w}{ }\PYG{esc}{\ensuremath{\forall}}\PYG{+w}{ }\PYG{n}{k}\PYG{o}{,}\PYG{+w}{ }\PYG{n}{k}\PYG{+w}{ }\PYG{n+nb+bp}{\PYGZlt{}}\PYG{+w}{ }\PYG{n}{gid}\PYG{+w}{ }\PYG{esc}{\ensuremath{\to}}\PYG{+w}{ }\PYG{n}{out}\PYG{o}{[}\PYG{n}{k}\PYG{o}{]}\PYG{n+nb+bp}{!}\PYG{+w}{ }\PYG{n+nb+bp}{=}\PYG{+w}{ }\PYG{n}{spec}\PYG{+w}{ }\PYG{n}{x}\PYG{o}{[}\PYG{n}{k}\PYG{o}{]}\PYG{n+nb+bp}{!}\PYG{o}{)}\PYG{+w}{ }\PYG{o}{:}
\PYG{+w}{    }\PYG{esc}{\ensuremath{\forall}}\PYG{+w}{ }\PYG{n}{k}\PYG{o}{,}\PYG{+w}{ }\PYG{n}{k}\PYG{+w}{ }\PYG{n+nb+bp}{\PYGZlt{}}\PYG{+w}{ }\PYG{n}{gid}\PYG{+w}{ }\PYG{n+nb+bp}{+}\PYG{+w}{ }\PYG{l+m+mi}{1}\PYG{+w}{ }\PYG{esc}{\ensuremath{\to}}\PYG{+w}{ }\PYG{o}{(}\PYG{n}{out.set}\PYG{n+nb+bp}{!}\PYG{+w}{ }\PYG{n}{gid}\PYG{+w}{ }\PYG{n}{v}\PYG{o}{)}\PYG{o}{[}\PYG{n}{k}\PYG{o}{]}\PYG{n+nb+bp}{!}\PYG{+w}{ }\PYG{n+nb+bp}{=}\PYG{+w}{ }\PYG{n}{spec}\PYG{+w}{ }\PYG{n}{x}\PYG{o}{[}\PYG{n}{k}\PYG{o}{]}\PYG{n+nb+bp}{!}
\end{Verbatim}

Only the reference \texttt{spec} changes between problems; the statement and its proof are otherwise identical, and it was re-proved in over fifty of the problem-runs. Two further obligations recurred in the same way: the reduction of a matrix-multiply entry to its sequential accumulation loop, across the matmul and gemm families, and a family of bridges relating a piecewise activation written as a nested conditional in the kernel to the clamp or maximum the specification denotes, across the rectifier variants and the fused epilogues.

Each of these is problem-independent, yet under the current design it consumes part of the Prover's fixed per-round budget on every problem, budget that, on the hardest problems, is the difference between a completed proof and an exhausted attempt. This points less to a bespoke lemma library than to a need for stronger automation. Developments that made facts of this kind available to \texttt{loom\_solve} directly, whether through a maturing TorchLean with reusable results about its tensor and reduction operators, or through richer \texttt{Float32} reasoning in the underlying floating-point libraries, would let the SMT backend discharge many of these obligations without any interactive step at all. Such automation would shift what the benchmark measures from proof-engineering stamina towards the faithfulness of the generated kernel, and would narrow the gap between stronger and weaker models, since much of the observed gap on the reduction and gemm families reflects the weaker model's inability to reconstruct these standard facts within the budget rather than a difference in the kernels it wrote. Improving the automation available for reasoning about these kernels is, in our view, one of the more valuable directions for making this style of verification practical.



\newpage
\section*{NeurIPS Paper Checklist}

\begin{enumerate}

\item {\bf Claims}
    \item[] Question: Do the main claims made in the abstract and introduction accurately reflect the paper's contributions and scope?
    \item[] Answer: \answerYes{} 
    \item[] Justification: The contributions stated in the abstract and Section~\ref{sec:Introduction} (agentic semantic and safety verification of LLM-generated CUDA kernels over \texttt{Float32}, and the identification of semantic failures that pass conventional tests) correspond directly to the results in Section~\ref{sec:Results} (Table~\ref{tab:aggregate}, Figure~\ref{fig:combined}) and the case studies in Section~\ref{sec:CaseStudy}, with the underlying assumptions (explicit preconditions, opaque transcendental functions) stated in Section~\ref{sec:ArchitectureAndSetup}.

\item {\bf Limitations}
    \item[] Question: Does the paper discuss the limitations of the work performed by the authors?
    \item[] Answer: \answerYes{} 
    \item[] Justification: Section~\ref{sec:Conclusion} contains a dedicated Limitations paragraph covering the restricted CUDA fragment handled by the deterministic compiler, the pointwise-agreement assumption required to transfer results over opaque transcendental functions to executed code, and the single-campaign scope of the evaluation.

\item {\bf Theory assumptions and proofs}
    \item[] Question: For each theoretical result, does the paper provide the full set of assumptions and a complete (and correct) proof?
    \item[] Answer: \answerNA{} 
    \item[] Justification: The paper does not include any theoretical results. 

    \item {\bf Experimental result reproducibility}
    \item[] Question: Does the paper fully disclose all the information needed to reproduce the main experimental results of the paper to the extent that it affects the main claims and/or conclusions of the paper (regardless of whether the code and data are provided or not)?
    \item[] Answer: \answerYes{} 
    \item[] Justification: Section~\ref{sec:ArchitectureAndSetup} specifies every pipeline component and Section~\ref{sec:Results} the full protocol: the 30 selected KernelBench problems, the five-valid-round loop, the 300-second per-round prover budget, the exact model variants, the feedback mechanism, and the orchestration harness; all rounds, kernels, and outcomes were logged.

\item {\bf Open access to data and code}
    \item[] Question: Does the paper provide open access to the data and code, with sufficient instructions to faithfully reproduce the main experimental results, as described in supplemental material?
    \item[] Answer: \answerYes{} 
    \item[] Justification: The complete pipeline (campaign controller, deterministic CUDA-to-Lean translator, specification modules, agent prompts, and vendored Lean dependencies) is included as supplementary material, with a README covering requirements, model-access prerequisites, setup, build, tests, and the exact command reproducing the campaign. The code accompanies the submission for review purposes and is not otherwise publicly released.

\item {\bf Experimental setting/details}
    \item[] Question: Does the paper specify all the training and test details (e.g., data splits, hyperparameters, how they were chosen, type of optimizer) necessary to understand the results?
    \item[] Answer: \answerYes{} 
    \item[] Justification: No models were trained. Section~\ref{sec:Results} details the complete experimental protocol, including problem selection, the round structure, agent budgets, the feedback loop, and the underlying models.

\item {\bf Experiment statistical significance}
    \item[] Question: Does the paper report error bars suitably and correctly defined or other appropriate information about the statistical significance of the experiments?
    \item[] Answer: \answerNo{} 
    \item[] Justification: Results are pass/fail counts from a single campaign per model over 30 problems (Table~\ref{tab:aggregate}); repeated campaigns to characterise run-to-run variance in agent behaviour were not run for cost reasons, as stated in the Limitations paragraph of Section~\ref{sec:Conclusion}.

\item {\bf Experiments compute resources}
    \item[] Question: For each experiment, does the paper provide sufficient information on the computer resources (type of compute workers, memory, time of execution) needed to reproduce the experiments?
    \item[] Answer: \answerYes{} 
    \item[] Justification: Section~\ref{sec:Results} states that every verification stage runs CPU-only on a single workstation with the LLMs accessed through their public APIs, and reports the total pipeline wall-clock time per model (roughly 14 hours over 131 rounds for GPT-5-fast and 9 hours over 102 rounds for Opus-5-fast); no further compute was required by the project.
    
\item {\bf Code of ethics}
    \item[] Question: Does the research conducted in the paper conform, in every respect, with the NeurIPS Code of Ethics \url{https://neurips.cc/public/EthicsGuidelines}?
    \item[] Answer: \answerYes{} 
    \item[] Justification: The research verifies machine-generated GPU kernels; it involves no human subjects, personal data, or high-risk models or datasets.

\item {\bf Broader impacts}
    \item[] Question: Does the paper discuss both potential positive societal impacts and negative societal impacts of the work performed?
    \item[] Answer: \answerYes{} 
    \item[] Justification: Section~\ref{sec:Conclusion} contains a Broader impact paragraph: the positive impact is a route to trusting LLM-generated code in production systems by catching silent numerical defects that testing misses; the risk is over-trust in certificates whose specifications or assumptions do not hold in deployment, mitigated by keeping every assumption explicit and stating when guarantees transfer to executed code.
    
\item {\bf Safeguards}
    \item[] Question: Does the paper describe safeguards that have been put in place for responsible release of data or models that have a high risk for misuse (e.g., pre-trained language models, image generators, or scraped datasets)?
    \item[] Answer: \answerNA{} 
    \item[] Justification: The paper releases no data or models that pose a risk of misuse.

\item {\bf Licenses for existing assets}
    \item[] Question: Are the creators or original owners of assets (e.g., code, data, models), used in the paper, properly credited and are the license and terms of use explicitly mentioned and properly respected?
    \item[] Answer: \answerYes{} 
    \item[] Justification: All existing assets --- KernelBench, TorchLean, Velvet, VerCors, Mathlib,  cvc5, \emph{torch.fx}, and OpenCode --- are credited by citation in Sections~\ref{sec:Background} and~\ref{sec:ArchitectureAndSetup} and are used in accordance with their open-source licenses.

\item {\bf New assets}
    \item[] Question: Are new assets introduced in the paper well documented and is the documentation provided alongside the assets?
    \item[] Answer: \answerYes{} 
    \item[] Justification: The new assets (the campaign controller, the CUDA-to-Lean translator, and the 30 machine-checked specification modules) are provided in the supplementary material, documented by the repository README and the campaign runbook (CAMPAIGN.md); they are provided for review rather than as a public release.

\item {\bf Crowdsourcing and research with human subjects}
    \item[] Question: For crowdsourcing experiments and research with human subjects, does the paper include the full text of instructions given to participants and screenshots, if applicable, as well as details about compensation (if any)? 
    \item[] Answer: \answerNA{} 
    \item[] Justification: The paper involves neither crowdsourcing nor research with human subjects.

\item {\bf Institutional review board (IRB) approvals or equivalent for research with human subjects}
    \item[] Question: Does the paper describe potential risks incurred by study participants, whether such risks were disclosed to the subjects, and whether Institutional Review Board (IRB) approvals (or an equivalent approval/review based on the requirements of your country or institution) were obtained?
    \item[] Answer: \answerNA{} 
    \item[] Justification: The paper involves neither crowdsourcing nor research with human subjects.

\item {\bf Declaration of LLM usage}
    \item[] Question: Does the paper describe the usage of LLMs if it is an important, original, or non-standard component of the core methods in this research? Note that if the LLM is used only for writing, editing, or formatting purposes and does \emph{not} impact the core methodology, scientific rigor, or originality of the research, declaration is not required.
    \item[] Answer: \answerYes{} 
    \item[] Justification: LLM agents are the object of study rather than an incidental tool: Sections~\ref{sec:ArchitectureAndSetup} and~\ref{sec:Results} fully describe how the Generator and Prover agents use GPT-5.6 Terra and Claude Opus 5 (fast variants), including their roles, budgets, and the verification feedback they receive.

\end{enumerate}

\end{document}